# Overview and Validation of the CEM and LAQGSM Event Generators for MCNP6, MCNPX, and MARS15

S. G. Mashnik

Los Alamos National Laboratory, Los Alamos, NM 87545, USA

**Abstract**

*A brief description of the IntraNuclear cascade, preequilibrium, evaporation, fission, coalescence, and Fermi breakup models used by the last versions of our CEM and LAQGSM event generators is presented, with a focus on our latest development of all these models. The recently developed "S" and "G" versions of our codes, that consider multifragmentation of nuclei formed after the preequilibrium stage of reactions when their excitation energy is above 2A MeV using the Statistical Multifragmentation Model (SMM) code by Botvina et al. ("S" stands for SMM) and the fission-like binary-decay model GEMINI by Charity ("G" stands for GEMINI), respectively, are overviewed as well. Examples of benchmarking our models against a large variety of experimental data on particle-particle, particle nucleus, and nucleus-nucleus reactions at energies from ≈ 10 MeV/A to ≈ 1 TeV/A, involving some very recent measurements of interest to accelerator radiation induced activation, are discussed.*

## 1. Introduction

Following an increased interest in intermediate- and high-energy nuclear data in relation to such projects as the Accelerator Transmutation of nuclear Wastes (ATW), Accelerator Production of Tritium (APT), Spallation Neutron Source (SNS), Rare Isotope Accelerator (RIA), Proton Radiography (PRAD) as a radiographic probe for the Advanced Hydro-test Facility, NASA needs, and others, the US Department of Energy has supported during the last decade our work on the development of improved versions of the Cascade-Exciton Model (CEM) and of the Los Alamos version of the Quark Gluon String Model (LAQGSM) which has led to our intermediate- and high-energy event generators CEM03.03 and LAQGSM03.03, respectively, and their modifications described below.

The main focus of our workshop is neutron- and accelerator-induced activation; radioactive waste generation, transmutation and burnup; medical isotope production; and radionuclide production in accelerator-driven physics facilities involving mainly nucleon-induced reactions at energies bellow several GeV. Such reactions are usually described well enough by our intermediate-energy code CEM03.03 without a need to use our high-energy code LAQGSM03.03. This is why we focus below mostly on CEM03.03 and its modifications. However, CEM does not consider the so-called "trawling" effect (depletion of target nucleons during a cascade), therefore does not describe well reactions on very light nuclei like C at incident energies above about 1 GeV. Therefore, in transport codes that use both CEM and our high-energy code LAQGSM as event generators, we recommend simulating nuclear reactions with CEM at incident energies up to about 1 GeV for light nuclei like C and up to about 5 GeV for actinide nuclei, and to switch simulations by LAQGSM, which does consider the "trawling" effect, at higher energies of transported particles. This is the reason we have included in the present talk a brief description of LAQGSM as well.

The Cascade-Exciton Model (CEM) of nuclear reactions was proposed almost 30 years ago at the Laboratory of Theoretical Physics, JINR, Dubna, USSR by Gudima, Mashnik, and Toneev [1]. It is based on the standard (non time-dependent) Dubna IntraNuclear Cascade (INC) [2, 3] and the Modified Exciton Model (MEM) [4, 5]. The code LAQGSM03.03 is the latest modification [6] of LAQGSM [7], which in its turn is an improvement of the Quark-Gluon String Model (QGSM) [8]. It describes reactions induced by both particles and nuclei at incident energies up to about 1 TeV/nucleon.

The basic version of both our CEM and LAQGSM event generators is the so-called "03.01" version, namely CEM03.01 [9] and LAQGSM03.01 [10]. The CEM03.01 code calculates nuclear reactions induced by nucleons, pions, and photons. It assumes that the reactions occur generally in three stages. The first stage is the INC, in which primary particles can be re-scattered and produce secondary particles several times prior to absorption by, or escape from the nucleus. When the cascade stage of a reaction is completed, CEM03.01 uses the coalescence model to "create" high-energy d, t, $^3$He, and $^4$He by final-state interactions among emitted cascade nucleons, already outside of the target. The emission of the cascade particles determines the particle-hole configuration, $Z$, $A$, and the excitation energy that is the starting point for the second, preequilibrium stage of the reaction. The subsequent relaxation of the nuclear excitation is treated in terms of an improved version of the modified exciton model of preequilibrium decay followed by the equilibrium evaporation/fission stage of the reaction. Generally, all four components may contribute to experimentally measured particle spectra and other distributions. But if the residual nuclei after the INC have atomic numbers with $A < 13$, CEM03.01 uses the Fermi breakup model to calculate their further disintegration instead of using the preequilibrium and evaporation models. Fermi breakup is much faster to calculate and gives results very similar to the continuation of the more detailed models to much lighter nuclei. LAQGSM03.01 also describes nuclear reactions, generally, as a three-stage process: INC, followed by preequilibrium emission of particles during the equilibration of the excited residual nuclei formed after the INC, followed by evaporation of particles from or fission of the compound nuclei. LAQGSM was developed with a primary focus on describing reactions induced by nuclei, as well as induced by most elementary particles, at high energies, up to about 1 TeV/nucleon. The INC of LAQGSM is completely different from the one in CEM. LAQGSM03.01 also considers Fermi breakup of nuclei with $A < 13$ produced after the cascade, and the coalescence model to "produce" high-energy d, t, $^3$He, and $^4$He from nucleons emitted during the INC.

The main difference of the following, so-called "03.02" versions of CEM and LAQGSM from the basic "03.01" versions is that the latter use the Fermi breakup model to calculate the disintegration of light nuclei instead of using the preequilibrium and evaporation models only after the INC, when the excited nuclei after the INC have a mass number $A < 13$, but do not use the Fermi breakup model at the preequilibrium, evaporation, and fission stages, when, due to emission of preequilibrium particles or due to evaporation or to a very asymmetric fission, we get an excited nucleus or a fission fragment with $A < 13$. This problem was solved in the 03.02 versions of CEM and LAQGSM, where the Fermi breakup model is used at any stage of a reaction, when we get an excited nucleus with $A < 13$.

In addition, the routines that describe the Fermi breakup model in the basic 03.01 version of our codes were written many years ago in the group of the Late Prof. Barashenkov at JINR, Dubna and are far from being perfect, though they are quite reliable and are still used currently without any changes in some transport codes. First, these routines allow in rare cases production of some light unstable fragments like $^5$He, $^5$Li, $^8$Be, $^9$B, etc., as a result of a breakup of some light excited nuclei. Second, these routines allowed in some very rare cases even production of "neutron stars" (or "proton stars"), *i.e.*, residual "nuclei" produced via Fermi breakup that consist of only neutrons (or only protons). Lastly, in some very rare cases, these routines could even crash the code, due to cases of 0/0. All these problems of the Fermi breakup model routines are addressed and solved in the 03.02 version of our codes. Several bugs are also fixed in 03.02 in comparison with its predecessor. On the whole, the 03.02 versions describe nuclear reactions on intermediate and light nuclei and production of fragments heavier than $^4$He from heavy targets much better than their predecessors, almost do not produce any unstable unphysical final products, and are free of the fixed bugs.

However, even after solving these problems and after implementing the improved Fermi breakup model into CEM03.02 and LAQGSM03.02, in some very rare cases, our event generators still could produce some unstable products via very asymmetric fission, when the excitation energy of such fragments was below 3 MeV and they were not checked and not disintegrated with the Fermi breakup model (see more details in [11]). This problem was addressed in the 03.03 versions of our codes, where we force such unstable products to disintegrate via Fermi breakup independently of their excitation energy. Several more bugs were fixed on the 03.03 version as well.

Details, examples of results, and useful references to different versions of our codes may be found in our recent lecture [11]. Many people participated in the CEM and LAQGSM code development over their almost 40-year history. Current contributors are S. G. Mashnik, K. K. Gudima, A. J. Sierk, R. E. Prael, M. I. Baznat, and N. V. Mokhov.

## 2. The Intranuclear Cascade Mechanism

The INC approach is based on the ideas of Heisenberg and Serber, who regarded intranuclear cascades as a series of successive quasi-free collisions of the fast primary particle with the individual nucleons of the nucleus. Basic assumptions of and conditions for INC applicability may be found in [11]. Comprehensive details and useful references are published in [2, 3].

### 2.1. The INC of CEM03.03

The intranuclear cascade model in CEM03.03 is based on the standard (non-time-dependent) version of the Dubna cascade model [2, 3]. All the cascade calculations are carried out in a three-dimensional geometry. The nuclear matter density $\rho(r)$ is described by a Fermi distribution with two parameters taken from the analysis of electron-nucleus scattering. For simplicity, the target nucleus is divided by concentric spheres into seven zones in which the nuclear density is considered to be constant. The energy spectrum of the target nucleons is estimated in the perfect Fermi-gas approximation. The influence of intranuclear nucleons on the incoming projectile is taken into account by adding to its laboratory kinetic energy an effective real potential, as well as by considering the Pauli principle which forbids a number of intranuclear collisions and effectively increases the mean free path of cascade particles inside the target. The interaction of the incident particle with the nucleus is approximated as a series of successive quasi-free collisions of the fast cascade particles (*N, π, or γ*) with intranuclear nucleons.

The integral cross sections for the free *NN*, *πN*, and *γN* interactions are approximated in the Dubna INC model [2,3] using a special algorithm of interpolation/extrapolation through a number of picked points, mapping as well as possible the experimental data. This was done very accurately by the group of Prof. Barashenkov using all experimental data available at that time, about 40 years ago [12]. Currently the experimental data on cross sections is much more complete than at that time; therefore we have revised the approximations of all the integral elementary cross sections used CEM.

The kinematics of two-body elementary interactions and absorption of photons and pions by a pair of nucleons is completely defined by a given direction of emission of one of the secondary particles. The cosine of the angle of emission of secondary particles in the c.m. system is calculated by the Dubna INC with approximations based on available experimental data. For elementary interactions with more than two particles in the final state, the Dubna INC uses the statistical model to simulate the angles and energies of products (see details in [2]).

For the improved version of the INC in CEM03.01, we use currently available experimental data and recently published systematics proposed by other authors and have developed new approximations for angular and energy distributions of particles produced in nucleon-nucleon and photon-proton interactions. In addition, we have incorporated in all new versions of CEM a possibility to normalize the final results to systematics based on available experimental reaction cross sections. The condition for the transition from the INC stage of a reaction to preequilibrium was changed; on the whole, the INC stage in CEM03.01 is longer while the preequilibrium stage is shorter in comparison with previous versions. We have incorporated real binding energies for nucleons in the cascade instead of the approximation of a constant separation energy of 7 MeV used in the initial versions of the CEM and have imposed momentum-energy conservation for each simulated even (provided only "on the average" by the initial versions). We also changed and improved the algorithms of many INC routines and almost all INC routines were rewritten, which speeded up the code significantly; some preexisting bugs were fixed. Details, examples of results, and references to this portion of our work may be found in [11].

*2.2. The INC of LAQGSM03.03*

The INC of LAQGSM03.03 is described with a recently improved version [6, 10, 13] of the time-dependent intranuclear cascade model developed initially at JINR in Dubna, often referred to in the literature as the Dubna intranuclear Cascade Model, DCM (see [14] and references therein). The DCM models interactions of fast cascade particles ("participants") with nucleon spectators of both the target and projectile nuclei and includes as well interactions of two participants (cascade particles). It uses experimental cross sections at energies below 4.5 GeV/nucleon, or those calculated by the Quark-Gluon String Model [8, 15] at higher energies to simulate angular and energy distributions of cascade particles, and also considers the Pauli Exclusion Principle.

In contrast to the CEM03.01 version of the INC described above, DCM uses a continuous nuclear density distribution (instead of the approximation of several concentric zones, where inside each the nuclear density is considered to be constant); therefore, it does not need to consider refraction and reflection of cascade particles inside or on the border of a nucleus. It also keeps track of the time of an intranuclear collision and of the depletion of the nuclear density during the development of the cascade (the so-called "trawling effect" mentioned above) and takes into account the hadron formation time.

All the new approximations developed recently for the INC of CEM03.01 to describe total cross sections and elementary energy and angular distributions of secondary particles from hadron-hadron interactions have been incorporated also in the INC of LAQGSM [10]. Then, a new high-energy photonuclear reaction model based on the of the event generators for $\gamma p$ and $\gamma n$ reactions from the Moscow INC [16] kindly provided us by Dr. Igor Pshenichnov, and on the latest photonuclear version of CEM [17] was developed and incorporated into the INC of LAQGSM, which allows us to calculate reactions induced by photons with energies of up to tens of GeV. In the latest version of LAQGSM, LAQGSM03.03 [6], the INC was modified for a better description of nuclear reactions at very high energies (above 20 GeV/nucleon). Finally, the algorithms of many LAQGSM INC routines were changed and some INC routines were rewritten, which speeded up the code significantly; some preexisting bugs in the DCM were fixed; many useful comments were added. Details, examples of results, and references to this portion of our work may be found in [11].

## 3. The Coalescence Model

When the cascade stage of a reaction is completed, CEM03.0x and LAQGSM03.0x use the coalescence model described in Refs. [14] to "create" high-energy d, t, $^3$He, and $^4$He by final-state interactions among emitted cascade nucleons, already outside of the target nucleus. In contrast to most other coalescence models for heavy-ion-induced reactions, where complex-particle spectra are estimated simply by convolving the measured or calculated inclusive spectra of nucleons with corresponding fitted coefficients, CEM03.0x and LAQGSM03.0x use in their simulations of particle coalescence real information about all emitted cascade nucleons and do not use integrated spectra. We assume that all the cascade nucleons having differences in their momenta smaller than $p_c$ and the correct isotopic content form an appropriate composite particle. The coalescence radii $p_c$ were fitted for each composite particle in Ref. [14] to describe available data for the reaction Ne+U at 1.04 GeV/nucleon, but the fitted values turned out to be quite universal and were subsequently found to describe high-energy complex-particle production satisfactorily for a variety of reactions induced both by particles and nuclei at incident energies up to about 200 GeV/nucleon, when describing nuclear reactions with different versions of LAQGSM [11] or with its predecessor, the Quark-Gluon String Model (QGSM) [8]. These parameters are:

$$p_c(d) = 90 \text{ MeV/c}; \quad p_c(t) = p_c(^3\text{He}) = 108 \text{ MeV/c}; \quad p_c(^4\text{He}) = 115 \text{ MeV/c}. \tag{1}$$

As the INC of CEM03.0x is different from those of LAQGSM or QGSM, it is natural to expect different best values for $p_c$ as well. Our recent studies show that the values of parameters $p_c$ defined by Eq. (1) are also good for CEM03.01 for projectile particles with kinetic energies $T_0$ lower than 300 MeV and equal to or above 1 GeV. For incident energies in the interval 300 MeV $< T_0 \leq$ 1 GeV, a better overall agreement with the available experimental data is obtained by using values of $p_c$ equal to 150, 175, and 175 MeV/c for d, t ($^3$He), and $^4$He, respectively. These values of $p_c$ are fixed as defaults in CEM03.01. If several cascade nucleons are chosen to coalesce into composite particles, they are removed from the distributions of nucleons and do not contribute further to such nucleon characteristics as spectra, multiplicities, *etc*.

In comparison with the initial version [14], in CEM03.0x and LAQGSM03.0x, several coalescence routines have been changed/deleted and have been tested against a large variety of measured data on nucleon- and nucleus-induced reactions at different incident energies.

## 4. Preequilibrium Reactions

The subsequent preequilibrium interaction stage of nuclear reactions is considered by our current CEM and LAQGSM in the framework of the latest version of the Modified Exciton Model (MEM) [4, 5] as implemented in CEM03.01 [9]. At the preequilibrium stage of a reaction, we take into account all possible nuclear transitions changing the number of excitons $n$ with $\Delta n = +2, -2$, and 0, as well as all possible multiple subsequent emissions of n, p, d, t, $^3$He, and $^4$He. The corresponding system of master equations describing the behavior of a nucleus at the preequilibrium stage is solved by the Monte-Carlo technique [1].

CEM considers the possibility of fast d, t, $^3$He, and $^4$He emission at the preequilibrium stage of a reaction in addition to the emission of nucleons. We assume that in the course of a reaction $p_j$ excited nucleons (excitons) are able to condense with probability $\gamma_j$ forming a complex particle which can be emitted during the preequilibrium state. The "condensation" probability $\gamma_j$ is estimated as the overlap integral of the wave function of independent nucleons with that of the complex particle (see details in [1])

$$\gamma_j = p_j^3 (V_j/V)^{p_j-1} = p_j^3 (p_j/A)^{p_j-1}. \tag{2}$$

This is a rather crude estimate. In the usual way the values $\gamma_j$ are taken from fitting the theoretical preequilibrium spectra to the experimental ones. In CEM03.0x, to improve the description of preequilibrium complex-particle emission, we estimate $\gamma_j$ by multiplying the estimate provided by Eq. (2) by an empirical coefficient $M_j (A,Z,T_0)$ whose values are fitted to available nucleon-induced experimental complex-particle spectra. We fix the fitted values in data commons of CEM03.0x and complement them with special routines for their interpolation outside the region covered by our fitting.

CEM and LAQGSM predict forward-peaked (in the laboratory system) angular distributions for preequilibrium particles. For instance, CEM03.0x assumes that a nuclear state with a given excitation energy $E^*$ should be specified not only by the exciton number $n$ but also by the momentum direction $\Omega$. This calculation scheme is easily realized by the Monte-Carlo technique [1]. It provides a good description of double differential spectra of preequilibrium nucleons and a not-so-good but still satisfactory description of complex-particle spectra from different types of nuclear reactions at incident energies from tens of MeV to several GeV. For incident energies below about 200 MeV, Kalbach [18] has developed a phenomenological systematics for preequilibrium-particle angular distributions by fitting available measured spectra of nucleons and complex particles. As the Kalbach systematics are based on measured spectra, they describe very well the double-differential spectra of preequilibrium particles and generally provide a better agreement of calculated preequilibrium complex-particle spectra with data than does the CEM approach [1]. This is why we have incorporated into CEM03.0x and LAQGSM03.0x the Kalbach systematics [18] to describe angular distributions of both preequilibrium nucleons and complex particles at incident energies up to 210 MeV. At higher energies, we use the CEM approach [1].

The standard version of the CEM [1] provides an overestimation of preequilibrium particle emission from different reactions we have analyzed (see more details in [19]). One way to solve this problem suggested in Ref. [19] is to change the criterion for the transition from the cascade stage to the preequilibrium one, as described in Section 2.1. Another easy way suggested in Ref. [19] to shorten the preequilibrium stage of a reaction is to arbitrarily allow only transitions that increase the number of excitons, $\Delta n = +2$, *i.e.*, only allow the evolution of a nucleus toward the compound nucleus. In this case, the time of the equilibration will be shorter and fewer preequilibrium particles will be emitted, leaving more excitation energy for the evaporation. This approach was used in the CEM2k [19] version of the CEM and it allowed us to describe much better the p+A reactions measured at GSI in inverse kinematics at energies around 1

GeV/nucleon. Nevertheless, the "never-come-back" approach seems unphysical, therefore we no longer use it. We now address the problem of emitting fewer preequilibrium particles in the CEM by following Veselsky [20]. We assume that the ratio of the number of quasi-particles (excitons) $n$ at each preequilibrium reaction stage to the number of excitons in the equilibrium configuration $n_{eq}$, corresponding to the same excitation energy, to be a crucial parameter for determining the probability of preequilibrium emission $P_{pre}$ (see details in [9, 11, 20]).

Algorithms of many preequilibrium routines are changed and almost all these routines are rewritten, which has speeded up the code significantly. Finally, some bugs were fixed [9, 11].

## 5. Evaporation

CEM03.01 and LAQGSM03.01 and their later versions use an extension of the Generalized Evaporation Model (GEM) code GEM2 by Furihata [21] after the preequilibrium stage of reactions to describe evaporation of nucleons, complex particles, and light fragments heavier than $^4$He (up to $^{28}$Mg) from excited compound nuclei and to describe their fission, if the compound nuclei are heavy enough to fission ($Z \geq 65$).

Note that when including evaporation of up to 66 types of particles in GEM2, its running time increases significantly compared to the case when evaporating only 6 types of particles, up to $^4$He. The major particles emitted from an excited nucleus are n, p, d, t, $^3$He, and $^4$He. For most cases, the total emission probability of particles heavier than α is negligible compared to those for the emission of light ejectiles. Our detailed study of different reactions shows that if we study only nucleon and complex-particle spectra or only spallation and fission products and are not interested in light fragments, we can consider evaporation of only 6 types of particles in GEM2 and save much time, getting results very close to the ones calculated with the more time consuming "66" option. In CEM03.01 and LAQGSM03.01, we have introduced an input parameter called **nevtype** that defines the number of types of particles to be considered at the evaporation stage. We recommend that users of CEM03.01 and LAQGSM03.01 use 66 for the value of the input parameter **nevtype** only when they are interested in all fragments heavier than $^4$He; otherwise, we recommend the default value of 6 for **nevtype**, saving computing time. Alternatively, users may choose intermediate values of **nevtype**, for example 9 if one wants to calculate the production of $^6$Li, or 14 for modeling the production of $^9$Be and lighter fragments and nucleons only, while still saving computing time compared to running the code with the maximum value of 66. A detailed description of GEM2 as incorporated into CEM and LAQGSM may be found in [9, 11].

## 6. Fission

The fission model used in GEM2 is based on Atchison's model [22], often referred in the literature as the Rutherford Appleton Laboratory (RAL) fission model, which is where Atchison developed it. The Atchison fission model is designed to describe only fission of nuclei with $Z \geq 70$. It assumes that fission competes only with neutron emission, i.e., from the widths $\Gamma_j$ of n, p, d, t, $^3$He, and $^4$He, the RAL code calculates the probability of evaporation of any particle. When a charged particle is selected to be evaporated, no fission competition is taken into account. When a neutron is selected to be evaporated, the code does not actually simulate its evaporation, instead it considers that fission may compete, and chooses either fission or evaporation of a neutron according to the fission probability $P_f$. This quantity is treated by the RAL code differently for the elements above and below $Z = 89$.

The mass-, charge-, and kinetic energy-distribution of fission fragments are simulated by RAL using approximations based on available experimental data (see details in [9, 11, 21, 22]).

For CEM03.01 and LAQGSM03.01, we modified slightly [23] GEM2. First, we fixed several observed uncertainties and small errors in the 2002 version of GEM2 Dr. Furihata kindly sent us. Then, we extended GEM2 to describe fission of lighter nuclei, down to $Z \geq 65$, and modified it [23] so that it provides a good description of fission cross sections when it is used after our INC and preequilibrium models: If we had merged GEM2 with the INC and preequilibrium-decay modules of CEM or of LAQGSM without any modifications, the new code would not describe correctly fission cross sections (and the yields of fission fragments). This is because Atchison fitted the parameters of his RAL fission model when it followed the Bertini INC [24] which differs from ours. In addition, Atchison did not model preequilibrium emission. Therefore, the distributions of fissioning nuclei in $A$, $Z$, and excitation energy $E^*$ simulated by Atchison differ significantly from the distributions we get; as a consequence, all the fission characteristics are also different. Furihata used GEM2 coupled either with the Bertini INC [24] or with the ISABEL [25] INC code, which also differs from our INC, and did not include preequilibrium particle emission. Therefore the distributions of fissioning nuclei simulated by Furihata differ from those in our simulations, so the parameters adjusted by Furihata to work well with her INC are not appropriate for us. To get a good description of fission cross sections (and fission-fragment yields) we have modified two parameters in GEM2 as used in CEM03.01 and LAQGSM03.01 (see more details in [9, 11, 23]).

## 7. The Fermi Breakup Model

After calculating the coalescence stage of a reaction, CEM03.01 and LAQGSM03.01 move to the description of the last slow stages of the interaction, namely to preequilibrium decay and evaporation, with a possible competition of fission. But as mentioned above, if the residual nuclei have atomic numbers with $A < 13$, CEM03.01 and

LAQGSM03.01 use the Fermi breakup model [26] to calculate their further disintegration instead of using the preequilibrium and evaporation models. The newer 03.02 versions of our codes use the Fermi breakup model also during the preequilibrium and/or evaporation stages of reactions, when the residual nucleus has an atomic number with $A < 13$. Finally, the latest 03.03 versions of our codes use the Fermi breakup model also to disintegrate the unstable fission fragments with $A < 13$ that can be produced in very rare cases of very asymmetric fission. All formulas and details of the algorithms used in the version of the Fermi breakup model developed in the group of the Late Prof. Barashenkov at Joint Institute for Nuclear Research (JINR), Dubna, Russia and used by our codes may be found in [27].

In comparison with its initial version as described in [27], the Fermi breakup model used in CEM03.02 and LAQGSM03.02 has been modified to decay the unstable light fragments that were produced by the original code. First, the initial routines allowed in rare cases production of some light unstable fragments like $^5$He, $^5$Li, $^8$Be, $^9$B, *etc.* as a result of a breakup of some light excited nuclei. Second, they allowed very rarely even production of "neutron stars" (or "proton stars"), *i.e.*, residual "nuclei" produced via Fermi breakup that consist of only neutrons (or only protons). Lastly, those routines could even crash the code, due to cases of division by 0. All these problems of the Fermi breakup model routines were addressed and solved in CEM03.02; the changes were then put in LAQGSM03.02. Several bugs are also fixed. However, even after solving these problems and after implementing the improved Fermi breakup model into CEM03.02 and LAQGSM03.02, our event generators still could produce some unstable products via very asymmetric fission, when the excitation energies of those fragments were below 3 MeV so they were not checked and disintegrated with the Fermi breakup model. This was the reason why a universal checking of all unstable light products has been incorporated into CEM03.03 and LAQGSM03.03. Such unstable products are forced to disintegrate via Fermi breakup independently of their excitation energy. The latest versions of the CEM03 and LAQGSM03 event generators do not produce any such unstable products.

## 8. "S" and "G" Versions of CEM03.01 and LAQGSM03.01

We have tested our CEM03.01 and LAQGSM03.01 codes and their later 03.02 and 03.03 versions against a large variety of experimental data on particle-particle, particle-nucleus, and nucleus-nucleus reactions at energies from ~10 MeV to ~1 TeV per nucleon (see, *e.g.*, [6, 9, 10, 11, 13, 17, 23] and references therein). As a rule, our codes are able to predict well a large variety of nuclear reactions. Therefore, they can be employed with confidence as reliable event generators in transport codes used as "working horses" for ADS and other applications. This is true only if we are not interested in the production of intermediate-mass fragments from not too heavy targets that do not fission from an old, "orthodox" point of view on fission, and do not produce such fragments via "conventional" fission. To be able to predict production of intermediate-mass fragments from not too heavy targets at such intermediate energies with our codes, we still need to solve some problems, just like other similar Monte-Carlo codes have to do.

An example of such a problem is presented in the low-energy parts of the Ni(p,x)$^{21}$Ne excitation functions shown in Fig. 1; another example is exercised in Fig. 2.

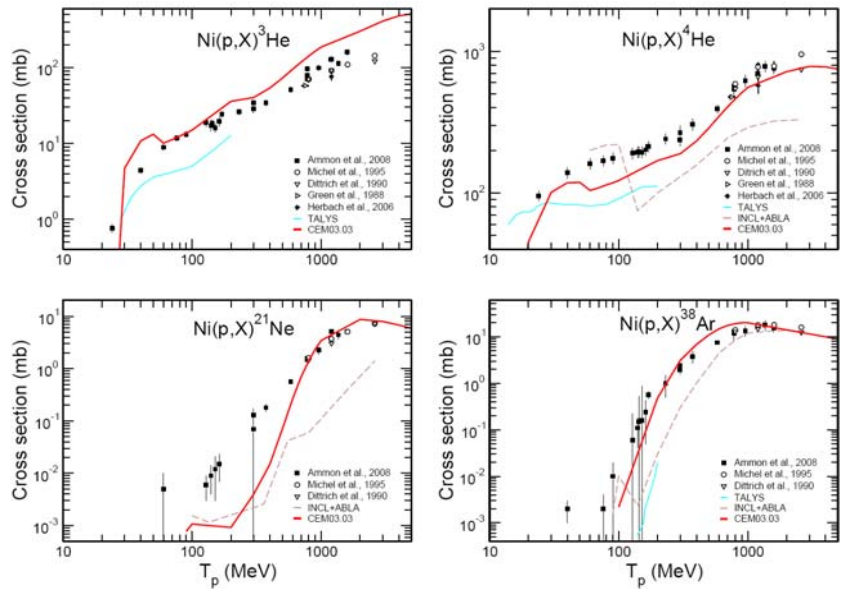

*Figure 1: Excitation functions for the production of $^3$He, $^4$He, $^{21}$Ne and $^{38}$Ar from p+Ni. Recent measurements by Ammon et al. [28] (filled squares) compared with our CEM03.03 results (red lines), with results by TALYS [29] (blue solid lines), and by INCL/ABLA [30] (dashed brown lines) from [28], and with previous measurements shown with different symbols, as indicated (see references to old data in [28]).*

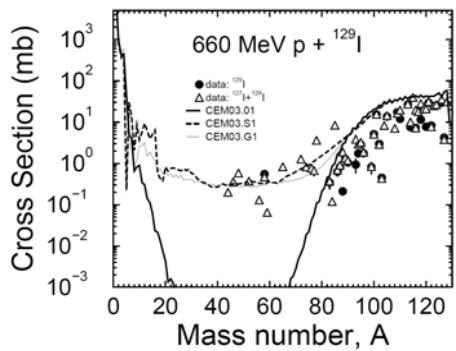

*Figure 2: Experimental [31] mass number distribution of the product yield from 660 MeV p + $^{129}I$ compared with calculations by CEM03.01 (thick solid line), CEM03.S1 (thick dashed line), and CEM03.G1 (solid thin line).*

We have addressed this problem in two different ways:

1) By implementing into CEM03.01 and LAQGSM03.01 the Statistical Multifragmentation Model (SMM) by Botvina *et al.* [32], to consider multifragmentation as a mode competitive to evaporation of particles and light fragments, when the excitation energy $E^*$ of a compound nucleus produced after the preequilibrium stage of a reaction is above $2 \times A$ MeV. This way, we have produced the "S" version of our codes ("S" stands for SMM), CEM03.S1 and LAQGSM03.S1.

CEM03.S1 and LAQGSM03.S1 are exactly the same as CEM03.01 and LAQGSM03.01, but consider also multifragmentation of excited nuclei produced after the preequilibrium stage of reactions, when their excitation energy is above $2 \times A$ MeV, using SMM. In SMM, within the total accessible phase space, a micro-canonical ensemble of all breakup configurations, composed of nucleons and excited intermediate-mass fragments governs the disassembly of the hot remnant. The probability of different channels is proportional to their statistical weight. Several different breakup partitions of the system are possible. When after the preequilibrium stage of a reaction $E^* > 2 \times A$ MeV and we "activate" SMM to calculate multifragmentation in the 03.S1 codes, the competitive evaporation process are calculated also with a version of the evaporation model by Botvina *et al.* from SMM [32], rather than using GEM2, as we do always in the standard 03.01 versions and in 03.S1 when $E^* \leq 2 \times A$ MeV and SMM is not invoked.

2) By replacing the Generalized Evaporation Model GEM2 by Furihata [21] used in CEM03.01 and LAQGSM03.01 with the fission-like binary-decay model GEMINI of Charity [33] which considers production of all possible fragments. This way, we have produced the "G" version of our codes ("G" stands for GEMINI), CEM03.G1 and LAQGSM03.G1.

CEM03.G1 and LAQGSM03.G1 are exactly the same as CEM03.01 and LAQGSM03.01, but use GEMINI instead of using GEM2. Within GEMINI, a special treatment based on the Hauser-Feshbach formalism is used to calculate emission of the lightest particles, from neutrons and protons up to beryllium isotopes. The formation of heavier nuclei than beryllium is modeled according to the transition-state formalism developed by Moretto. All asymmetric divisions of the decaying compound nuclei are considered in the calculation of the probability of successive binary-decay configurations. GEMINI is described in details in the original publications [33], therefore we do not elaborate here.

From Fig. 2, we see that both "S" and "G" versions reproduce almost equally well the yields of intermediate-mass fragments from these reactions that can not be described by the standard CEM03.01. From one point of view, this is an achievement, as, with the "S" and "G" versions, we are able to describe reasonably well the reactions where the standard 03.01 versions do not work. On the other hand, this also makes the situation more intricate, as from these (and other similar) results we can not choose easily between the "S" and "G" versions, that makes more difficult to determine the real mechanisms of such reactions. We think that for such intermediate-energy proton-induced reactions the contribution of multifragmentation to the production of heavy fragments should not be very significant due to the relatively low excitation energies involved. Such fragments are more likely to be produced via the fission-like binary decays modeled by GEMINI or/and via preequilibrium reactions. We conclude that it is impossible to make a correct choice between fission-like and fragmentation reaction mechanisms involved in these (or other similar) reactions merely by comparing model results with the measurements of only product cross sections; addressing this question will probably require analysis of two- or multi-particle correlation measurements (see more details in [11]).

## 9. Validation of CEM and LAQGSM

As mentioned above, we have tested our event generators against a large variety of experimental data on particle-particle, particle-nucleus, and nucleus-nucleus reactions at energies from ~10 MeV to ~1 TeV per nucleon. Many examples of a very good agreement between predictions by our codes and various measured reactions may be found, *e.g.*, in [6, 9, 10, 11, 13, 17, 23] and references therein. Below, we present a few examples of benchmarking our models against some very recent measurements, focusing on reactions where the calculations do not agree very well with the data, indicating that we have to improve further our models.

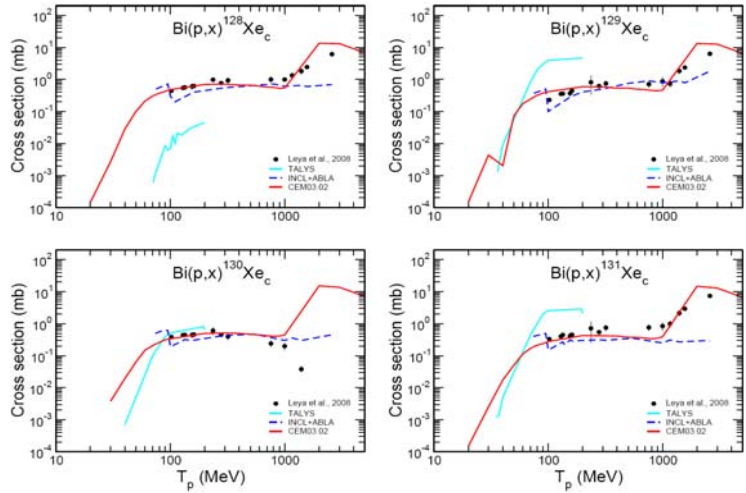

*Figure 3: Excitation functions for the cumulative production of $^{128}$Xe, $^{129}$Xe, $^{130}$Xe and $^{131}$Xe from p+Bi. Recent measurements by Leya et al. [34] (filled circles) are compared with our CEM03.02 results (red lines), with results by TALYS [29] (cyan solid lines), and by INCL/ABLA [30] (blue dashed lines) from [34].*

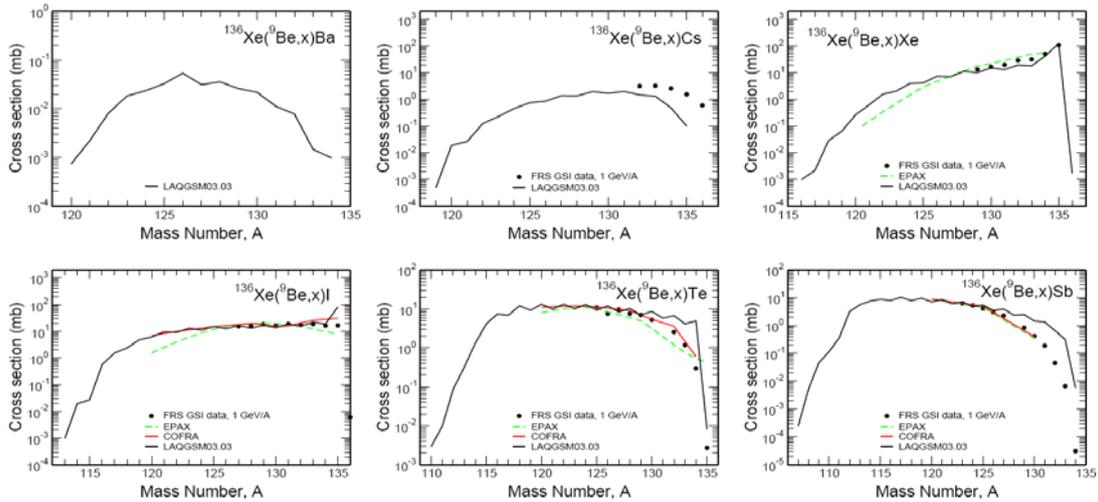

*Figure 4: Experimental (filled circles) [35] mass number distributions of Ba, Cs, Xe, I, Te, and Sb isotopes from 1 GeV/A $^{136}$Xe + $^{9}$Be compared with calculations by LAQGSM03.03 (black solid lines) and results by EPAX (green dashed lines), and COFRA (red sold lines) from [35]; for Ba, we show only predictions by LAQGSM03.03; no Ba isotopes were measured and no Ba, Cs, and Xe results by EPAX and COFRA are published in [35].*

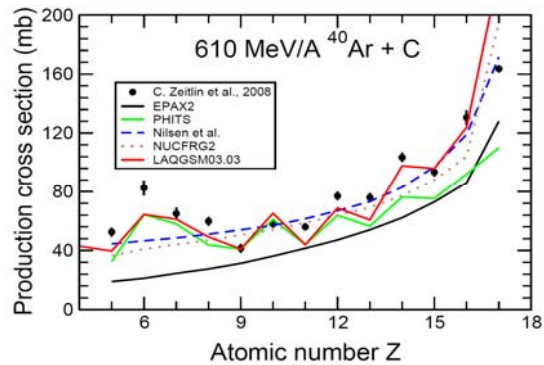

*Figure 5: Experimental (filled circles) [38] atomic number distribution of fragment cross sections for 610 MeV/A $^{40}$Ar*

+ $^{12}$C compared with our LAQGSM03.03 calculations (red solid line) and with results by EPAX2 [36] (black solid line), PHITS [39] (green solid line), a model by Nilsen et al. [40] (blue dashed line), and NUCFRG2 [41] (brown dotted line) from [38], as indicated.

We may observe some disagreements of our results with these new data, especially for the production of $^{130}$Xe from p + Bi (Fig. 3), Cs from $^{136}$X + $^{9}$Be (Fig. 4), and Cl from $^{40}$Ar + C (Fig. 5), indicating us that there are still problems to be solved in CEM and LAQGSM and we have to improve further our models to describe well these types of reactions. We see that TALYS [29], INCL/ABLA [30], EPAX [36], COFRA [37], NUCFRG2 [41], and the model by Nilson et al. [40] also encounter difficulties in reproducing these data, indicating that similar problems occur with other models as well.

## 10. Summary

Improved versions of the cascade-exciton model of nuclear reactions and of the Los Alamos quark-gluon string model have been developed recently at LANL and implemented in the codes CEM03.01 and LAQGSM03.01, and in their slightly corrected 03.02 and 03.03 versions. The recent versions of our codes describe quite well and much better than their predecessors a large variety of nuclear reactions of interest to ADS, as well as reactions at higher energies of interest to NASA, accelerator shielding, astrophysics, and other applications, up to ~1 TeV/nucleon.

We observe that codes developed for applications must not only describe reasonably well arbitrary reactions without any free parameters, but also not require too much computing time. Our own exercises on many different nuclear reactions calculated with MCNPX [42] and LAHET3 [43] transport codes using different event generators show that the current version of CEM, using **nevtype** = 6, is a little faster than the Bertini [24] and ISABEL [25] options for INC in these transport codes, and is several times faster than the INCL/ABLA [30] option. The computing time of LAQGSM for heavy-ion reactions was tested recently by Dr. Gomez (see [11] for more details) with MCNPX [42] in comparison with PHITS [39], on simulations of a 400 MeV/nucleon uranium beam on a 0.2-cm thick lithium target. He found that PHITS [39] using the Quantum Molecular Dynamics (QMD) model to simulate heavy-ion reactions requires 210 times more computing time than LAQGSM does, making it very difficult if not impossible to use in applications requiring fast results; LAQGSM does not have these problems, providing quite fast and reliable results.

Both our CEM and LAQGSM event generators still have some problems in a reliable description of some fragments from intermediate-energy nuclear reactions on medium-mass nuclei, just as other similar modern Monte-Carlo codes do. To address these problems, we developed "S" and "G" modifications of our codes, which consider multifragmentation of nuclei formed after the preequilibrium stage of reactions when their excitation energy is above 2×A MeV using the SMM code by Botvina et al. [32], and the fission-like binary-decay model GEMINI by Charity [33], respectively. The "S" and "G" versions of our codes allow us to describe many fragmentation reactions that we are not able to reproduce properly with the standard version of our codes. However, there are still some problems to be solved in understanding the "true" mechanisms of fragment productions from many reactions, therefore we consider the present "S" and "G" versions of our codes only as working modifications and they are not yet implemented as event generators into our transport codes, in contrast to our "standard" 03.01, 03.02, and 03.03 versions used as event generators.

The latest versions of our 03.01, 03.02, and 03.03 CEM and LAQGSM codes have been or are being incorporated as event generators into the transport codes MCNP6 [44], MCNPX [42], and MARS [45]. CEM03.01 was made available to the public via RSICC at Oak Ridge and NEA/OECD in Paris as the Code Package PSR-0532. We also plan to make LAQGSM03.01 and the latest 03.03 versions of our codes available to the public via RSICC and NEA/OECD in the future.

**Acknowledgements**

I thank the Organizers, and especially, Drs. Sabine Teichmann, Michael Wohlmuther, and Franz Gallmeier for inviting me to present this talk and for financial support of my trip to ARIA'08. This work was written and the CEM and LAQGSM codes were developed and tested under the auspices of the National Nuclear Security Administration of the U.S. Department of Energy at Los Alamos National Laboratory under Contract No. DE-AC52-06NA25396.